\documentclass[11pt]{article}
\usepackage{amsmath,amsfonts,latexsym,graphicx,amssymb}
\usepackage[dvips]{epsfig}
\date{}
\def\la{\langle\,}
\def\r{\,\rangle}
\newcommand{\eeq}{\end{eqnarray}}
\newcommand{\beq}{\begin{eqnarray}}

\newcommand{\mf}{\mathfrak{f}}
\def\con{{}_{\_\rule{-1pt}{0pt}\_}
\rule{-2pt}{0pt}\raise1.5pt\hbox{$\mid$}\hspace{2pt}}

\def\oper{{\mathchoice{\rm 1\mskip-4mu l}{\rm 1\mskip-4mu l}%
{\rm 1\mskip-4.5mu l}{\rm 1\mskip-5mu l}}}
\def\<{\langle}
\def\>{\rangle}
\title{\bf Quantum damped oscillator I: \\ dissipation and resonances}
\author{Dariusz Chru\'sci\'nski and Jacek Jurkowski\\
 Institute of Physics, Nicolaus Copernicus University\\
 ul. Grudzi\c{a}dzka 5/7, 87-100 Toru\'n, Poland}

\begin{document}

\maketitle

\begin{abstract}
Quantization  of a damped harmonic oscillator  leads to so called
Bateman's dual system. The corresponding Bateman's Hamiltonian,
being a self-adjoint operator, displays the discrete family of
complex eigenvalues. We show that they correspond to the poles of
energy eigenvectors and the corresponding resolvent operator when
continued to the complex energy plane. Therefore, the
corresponding generalized eigenvectors may be interpreted as
resonant states which are responsible for the irreversible quantum
dynamics of a damped harmonic oscillator.

\end{abstract}

\numberwithin{equation}{section}

\section{Introduction}
\setcounter{equation}{0}

The damped harmonic oscillator is one of the simplest quantum
systems displaying the dissipation of energy. Moreover, it is of
great physical importance and has found many applications
especially in quantum optics. For example it plays a central role
in the quantum theory of lasers and masers \cite{QO1,QO2,QO3}.

As is well known there is no room for the dissipative phenomena in
the standard Hilbert space formulation of Quantum Mechanics. The
Schr\"odinger equation defines one-parameter unitary group and
hence the quantum dynamics is perfectly time-reversible. The usual
approach to include dissipation is the quantum theory of open
systems \cite{Haake,Davies,Petruccione,Carmichael}. In this
approach the dynamics of a quantum system is no longer unitary but
it is defined by a semigroup of completely positive maps in the
space of density operators \cite{Alicki} (for recent reviews see
e.g. \cite{Benatti,Olkiewicz}).

There is, however, another way to describe dissipative quantum
systems based on  the old idea of Bateman \cite{Bat31}. Bateman
has shown that to apply the standard canonical formalism of
classical mechanics to dissipative and non-Hamiltonian systems,
one can double the numbers of degrees of freedom, so as to deal
with an effective isolated classical Hamiltonian system. The new
degrees of freedom may be assumed to represent a reservoir.
Applying this idea to damped harmonic oscillator one obtains a
pair of damped oscillators (so called Bateman's dual system): a
primary one and its time reversed image. The Bateman dual
Hamiltonian has been rediscovered by Morse and Feshbach
\cite{Mor53} and Bopp \cite{Bopp73} and the detailed quantum
mechanical analysis was performed by Feshbach and Tikochinski
\cite{FT}. The quantum Bateman system was then analyzed by many
authors (see the detailed historical review \cite{Dekker81} with
almost 600 references!).

Surprisingly, this system is still  worth to study and it shows
its new interesting features. Recently it was analyzed in
\cite{Vit92} in connection with quantum field theory and quantum
groups (see also \cite{Vit94,Vit95}). Different approach based on
the Chern-Simons theory was applied in  \cite{Ban02}. In a recent
paper \cite{Bla04} a damped oscillator was quantized by using
Feynman path integral formulation (see also \cite{Gerry}).
Moreover, the corresponding geometric phase was calculated and
found to be directly related to the ground-state energy of the
standard one-dimensional  linear harmonic oscillator. Bateman's
system has been also studied as a toy model for the recent
proposal by 't Hooft about deterministic quantum mechanics
\cite{t-Hooft1,t-Hooft2}.

In the present paper we propose a slightly different approach to
the Bateman system. The unusual feature of the Bateman Hamiltonian
is that being a self-adjoint operator it displays a family of
complex eigenvalues. We show that these eigenvalues correspond to
the poles of energy eigenvectors and the corresponding resolvent
operator when continued to the complex energy plane. The similar
analysis for the toy model of a quantum damped system was
performed in \cite{damp1,damp2}. Eigenvectors corresponding to the
poles of the resolvent are well known in the scattering theory as
resonant states \cite{RES-1,RES-2}. It shows that the appearance
of resonances is responsible for the dissipation in the Bateman
system. Obviously, the time evolution is perfectly reversible when
considered on the corresponding system Hilbert space ${\cal H} =
L^2( \mathbb{R}^2)$. It is given by the 1-parameter group of
unitary transformations $U(t)=e^{-i\widehat{H}t}$. It turns out
that there are two natural subspaces ${\cal S}_\pm \in {\cal H}$
such that $U(t)$ restricted to ${\cal S}_\pm$ defines only two
semigroups: $U(t\geq 0)$ on ${\cal S}_-$, and $U(t\leq 0)$ on
${\cal S}_+$. These two semigroups are related by the time
reversal operator $\cal T$ (see Section~\ref{Reversal}).

Our analysis is based on a new representation of the Bateman
Hamiltonian, cf.  Section~\ref{New-Rep}. This representation is
directly related to the old observation of Pontriagin \cite{Pontr}
(see Section~\ref{Sec-Pontr} for review) that any non-Hamiltonian
system of the form
\begin{equation}\label{}
    \dot{x}_k = X_k(x_1,\ldots,x_N) \ , \ \ \ \ k=1,2,\ldots,N\ ,
\end{equation}
may be treated as a Hamiltonian one in the extended phase-space
$(x_1,\ldots,x_N,p_1,\ldots,p_N)$ with the Hamiltonian
\begin{equation}\label{}
    H(x_1,\ldots,x_N,p_1,\ldots,p_N) = \sum_{k=1}^N p_k\,
    X_k(x_1,\ldots,x_N)\ .
\end{equation}
Note, that the above Hamiltonian has exactly the form considered
by 't Hooft \cite{t-Hooft1}.

From the mathematical point of view the natural language to
analyze the Bateman system is the so called rigged Hilbert space
approach to quantum mechanics \cite{RHS1,RHS2,Bohm-Gadella}. There
are two natural rigged Hilbert spaces, or Gel'fand triplets,
corresponding to subspaces ${\cal S}_\pm$.  We shall comment on
that in Section~\ref{CON}.

\section{Bateman Hamiltonian}
\label{sect-BHam} \setcounter{equation}{0}

The classical equation of motion for one-dimensional damped
oscillator with unit mass reads
\begin{equation}
 \label{damped_osc}
 \ddot{x}+2\gamma \dot{x}+ \kappa x \;=\; 0\ ,
\end{equation}
where $\gamma>0$ denotes the damping constant. Introducing
Bateman's dual system
\begin{equation}
 \label{dual}
 \ddot{y}-2\gamma \dot{y}+ \kappa y \;=\; 0\ ,
\end{equation}
one may derive booth equations from the following Lagrangian
\begin{equation}\label{Bateman_L}
    L(x,\dot{x},y,\dot{y})=\dot{x}\dot{y}
    -\kappa xy + \gamma(x\dot{y}-
    \dot{x}y)\ .
\end{equation}
Introducing  canonical momenta
\begin{equation}\label{}
    p_x = \dot{y} - \gamma y \,, \qquad p_y = \dot{x} +
    \gamma x \, ,
\end{equation}
one easily finds the corresponding Hamiltonian
\begin{equation}\label{BHam}
    H(x,y,p_x,p_y)=
    p_x p_y -\gamma(xp_x - yp_y)+ \omega^2 xy\ ,
\end{equation}
where
\begin{equation}\label{omega}
    \omega = \sqrt{\kappa - \gamma^2}\ .
\end{equation}
Throughout the paper we shall consider the underdamped case, i.e.
$\kappa > \gamma^2$.

Now, assuming symmetric Weyl ordering the canonical quantization is straightforward and leads to
the following self-adjoint  operator in the Hilbert space $L^2( \mathbb{R}^2,dxdy)$:
\begin{equation}\label{HHH}
\hat{H} = \hat{H}_0 + \hat{H}_I \ ,
\end{equation}
where
\begin{equation}\label{H0}
    \hat{H}_0 =
    \hat{p}_x \hat{p}_y + \omega^2 \hat{x}\hat{y}\ ,
\end{equation}
and
\begin{equation}\label{HI}
     \hat{H}_I = -\frac{\gamma}{2} \Big( (\hat{x}\hat{p}_x + \hat{p}_x\hat{x})
    - (\hat{y}\hat{p}_y + \hat{p}_y\hat{y})\Big)\ .
\end{equation}
Note, that
\begin{equation}\label{}
    [ \hat{H}_0,\hat{H}_I]=0\ .
\end{equation}
 Following Feshbach
and Tichochinsky \cite{FT} one  introduces annihilation and
creation operators
\begin{eqnarray}\label{FT1}
\hat{A} &=& \frac{1}{2\sqrt{\hbar\omega}}\Big[(\hat{p}_x + \hat{p}_y)-i\omega(\hat{x} + \hat{y})\Big]\,,\\
\hat{B} &=& \frac{1}{2\sqrt{\hbar\omega}}\Big[(\hat{p}_x -
\hat{p}_y)-i\omega(\hat{x} - \hat{y})\Big]\, . \label{FT2}
\end{eqnarray}
They satisfy the standard CCRs
\begin{equation}\label{CCR}
    [\hat{A},\hat{A}^\dag] =  [\hat{B},\hat{B}^\dag] = 1\ ,
\end{equation}
and all other commutators vanish. It turns out that the
transformed Hamiltonian is given by (\ref{HHH}) with
\begin{equation}\label{H_0_H_I}
    \hat{H}_0 = \hbar\omega(\hat{A}^\dagger \hat{A} - \hat{B}^\dagger
    \hat{B})\ , \qquad  \hat{H}_I = i\hbar\gamma (\hat{A}^\dagger
    \hat{B}^\dagger-\hat{A}\hat{B})\ .
\end{equation}
 It is easy to see \cite{FT,Vit92} that the dynamical symmetry
associated with the Bateman's Hamiltonian is that of $SU(1,1)$.
Indeed,   constructing the following  generators:
\begin{eqnarray}
\hat{J}_1 &=& \frac12\,(\hat{A}^\dagger \hat{B}^\dagger+\hat{A}\hat{B})\,,\\
\hat{J}_2 &=& \frac i2\,(\hat{A}^\dagger \hat{B}^\dagger-\hat{A}\hat{B})\,,\\
\hat{J}_3 &=& \frac12\,(\hat{A}^\dagger \hat{A} +
\hat{B}\hat{B}^\dagger)\,,
\end{eqnarray}
 one easily shows that  they satisfy $su(1,1)$ commutation
relations:
\begin{equation}\label{}
    [ \hat{J}_1,\hat{J}_2] = i \hat{J}_3\ , \ \ \ \ \ \
     [ \hat{J}_3,\hat{J}_2] = i \hat{J}_1\ , \ \ \ \ \ \
      [ \hat{J}_1,\hat{J}_3] = i \hat{J}_2\ .
\end{equation}
Moreover, the following operator
\begin{equation}\label{Casimir}
    \hat{J}_0 = \frac12\,(\hat{A}^\dagger \hat{A} - \hat{B}^\dagger \hat{B})\,,
\end{equation}
defines the corresponding $su(1,1)$ Casimir operator. One easily
shows that
\begin{equation}\label{}
    \hat{J}^2_0 = \frac 14 +  \hat{J}^2_3 -  \hat{J}^2_1 -
    \hat{J}^2_2 \ .
\end{equation}
It is therefore clear that the Hamiltonian (\ref{H_0_H_I}) can be
rewritten in terms of $su(1,1)$ generators as
\begin{equation}\label{}
\hat{H}_0 = 2\hbar\omega \hat{J}_0 \,, \qquad \hat{H}_I =
2\hbar\gamma \hat{J}_2\, .
\end{equation}
The algebraic structure arising in this approach enables one to
solve the corresponding eigenvalue problem. Let us define two mode
eigenvectors $|n_A,n_B\rangle$:\footnote{Mathematically oriented
reader would prefer
\begin{equation*}
(\hat{A}^\dagger \hat{A} \otimes \oper_B) |n_A,n_B\rangle = n_A
|n_A,n_B\rangle\, , \qquad (\oper_A \otimes \hat{B}^\dagger
\hat{B}) |n_A,n_B\rangle = n_B |n_A,n_B\rangle \,,
\end{equation*}
where $\oper_A$ ($\oper_B$) denotes the identity operator in
``$A$-sector'' (``$B$-sector").}
\begin{equation}\label{}
\hat{A}^\dagger \hat{A} |n_A,n_B\rangle = n_A |n_A,n_B\rangle \ ,
\hspace{1cm} \hat{B}^\dagger \hat{B} |n_A,n_B\rangle = n_B
|n_A,n_B\rangle \ .
\end{equation}
It is convenient to introduce
\begin{equation}\label{j-m}
    j = \frac 12 (n_A - n_B) \ , \hspace{1cm}  m =  \frac 12 (n_A
    + n_B)\ ,
\end{equation}
and to label the corresponding eigenvectors of $ \hat{J}_0 $ and $
\hat{J}_3$ by $|j,m\rangle$ rather than $|n_A,n_B\rangle$:
\begin{eqnarray}\label{}
\hat{J}_0|j,m\rangle &=& j|j,m\rangle\ , \\
\hat{J}_3|j,m\rangle &=& \left(m + \frac 12 \right) |j,m\rangle\ .
\end{eqnarray}
Clearly,
\begin{equation}\label{jm!!!}
    j = 0, \pm \frac 12, \pm 1, \pm \frac 32, \ldots , \hspace{1cm} m = |j|, |j| +1,
    |j| + 2, \ldots \ .
\end{equation}
 Finally, defining
\begin{equation}\label{psi-jm}
    |\psi^\pm_{jm}\rangle = \exp\left(\mp \frac \pi 2 \hat{J}_1
    \right) |jm\rangle\ ,
\end{equation}
one obtains
\begin{equation}\label{}
\hat{H}|\psi^\pm_{jm}\rangle = E^\pm_{jm} |\psi^\pm_{jm}\rangle\ ,
\end{equation}
with
\begin{equation}\label{FTspectrum}
  E^{\pm}_{jm} = 2\hbar\omega j \pm i\hbar\gamma(2m+1)\ .
\end{equation}
Let us emphasize that the eigenvectors corresponding to energies
(\ref{FTspectrum}) cannot be normalized and should be considered
as generalized eigenvectors not belonging to the Hilbert space of
the problem.

\section{Canonical  quantization of non-Hamiltonian systems}
\label{Sec-Pontr}
\setcounter{equation}{0}

As is well known any dynamical system may be regarded as a part of
a larger Hamiltonian system. Bateman's approach is based on adding
to the primary system a time reversed (dual)  copy. Together they
define a Hamiltonian system. There exists, however, a general
approach to canonical quantization of non-Hamiltonian systems based
on an old observation of Pontriagin \cite{Pontr}.
Suppose we are given an arbitrary non-Hamiltonian system described
by
\begin{equation}\label{dotx-X}
  \dot{\bf x} = {\bf X}({\bf x})\ ,
\end{equation}
where $\bf X$ is a vector field on some configuration space $Q$.
For simplicity assume that $Q \subset \mathbb{R}^N$, that is, the
system has $N$ degrees of freedom. This system may be lifted to
the Hamiltonian system on the phase space ${\cal P} = Q \times
\mathbb{R}^N$ as follows: one defines the Hamiltonian $
  H  : {\cal P} \longrightarrow \mathbb{R}$ by
\begin{equation}\label{H-xp}
  H({\bf x},{\bf p}) = {\bf p}\cdot {\bf X}({\bf x}) = \sum_{l=1}^N \, p_l X_l({\bf x})\ ,
\end{equation}
where $({\bf x},{\bf p}) = (x_1,\ldots,x_N,p_1,\ldots,p_N)$ denote
canonical coordinates on $\cal P$. The corresponding Hamilton
equations read as follows:
\begin{eqnarray}\label{HAM-1}
  \dot{x}_k &=& \{ x_k,H\} = X_k(x) \ , \\
  \dot{p}_k &=& \{ p_k,H\} = - \sum_{l=1}^N p_l \frac{\partial
  X_l(x)}{\partial x_k} \ ,
\end{eqnarray}
for $k=1,\ldots,N$. In the above formulae $\{\ , \ \}$ denotes the
standard Poisson bracket on $\cal P$
\begin{equation}\label{}
  \{F,G\} = \sum_{k=1}^N \left( \frac{\partial F}{\partial x_k}
   \frac{\partial G}{\partial p_k} - \frac{\partial G}{\partial x_k}
   \frac{\partial F}{\partial p_k}  \right) \ .
\end{equation}
Clearly, the formulae (\ref{HAM-1}) reproduce our initial
dynamical system (\ref{dotx-X}) on $Q$. The canonical quantization is now straightforward.
Assuming the symmetric Weyl ordering one obtains the following formula for the quantum Hamiltonian
\begin{equation}\label{}
    \hat{H}_{\rm quantum} = {\rm W} \left(  \sum_{l=1}^N \, p_l X_l({\bf x}) \right)\ ,
\end{equation}
where ${\rm W}(f)$ denotes the Wigner-Weyl transform of a
space-phase function $f=f({\bf x},{\bf p})$. Recall, that the
Wigner-Weyl transform of $f$ is defined as follows
\begin{equation}\label{WW}
    \hat{f} = {\rm W}(f) =  \int d\mbox{\boldmath $\sigma$} \int d\mbox{\boldmath $\tau$}\,
    \widetilde{f}(\mbox{\boldmath $\sigma$},\mbox{\boldmath $\tau$}) \exp\left\{ i \sum_{k=1}^N
    \left( \sigma_k \hat{x}_k + \tau_k \hat{p}_k\right) \right\} \ ,
\end{equation}
where $ \widetilde{f}(\mbox{\boldmath $\sigma$},\mbox{\boldmath $\tau$})$ denotes the Fourier
transform of $f({\bf x},{\bf p})$. Clearly, $\hat{H}_{\rm quantum}$ defines a Schr\"odinger system in
$L^2( \mathbb{R}^N,d{\bf x})$.

Consider now a  damped harmonic oscillator described by
\begin{equation*}\label{}
  \ddot{x} + 2\gamma \dot{x} + \kappa x = 0 \ .
\end{equation*}
The above 2nd order equation may be rewritten as a dynamical
system on $ \mathbb{R}^2$
\begin{eqnarray}\label{}
  \dot{x}_1 &=& - \gamma x_1 + \omega x_2 \ , \\
  \dot{x}_2 &=& - \gamma x_2 - \omega x_1\ ,
\end{eqnarray}
with $\omega$ defined in (\ref{omega}). Clearly this system is not
Hamiltonian if $\gamma \neq 0$. However, applying the above
Pontriagin procedure  one arrives at the Hamiltonian system on $
\mathbb{R}^4$ defined by the following damped harmonic oscillator
 Hamiltonian:
\begin{equation}\label{H-osc-d}
  H({\bf x},{\bf p}) = \omega( p_1x_2 - p_2x_1) - \gamma(p_1x_1 + p_2x_2)\ .
\end{equation}
The corresponding Hamilton equations of motion read
\begin{equation}\label{}
    \dot{\bf x} = \hat{F} {\bf x} \ , \hspace{1cm} \dot{\bf p} = -\hat{F}^{\rm T} {\bf
    p}\ ,
\end{equation}
where
\begin{equation}\label{}
    \hat{F} = \left( \begin{array}{cc} -\gamma & \omega \\ -\omega
    & -\gamma \end{array} \right)\ ,
\end{equation}
and $\hat{F}^{\rm T}$ denotes the transposition of $\hat{F}$. One
may ask what is the relation between Bateman's Hamiltonian
(\ref{BHam}) and that obtained via Pontriagin procedure
(\ref{H-osc-d}). Surprisingly they are related by the following
simple canonical transformation $(x,y,p_x,p_y) \longrightarrow (x_1,x_2,p_1,p_2)$:

\begin{eqnarray}
x_1 & = &   \frac{p_y}{\sqrt{\omega}} \ , \hspace{1.5cm} p_1\, = \,-\sqrt{\omega}\, y \\
x_2 & = &  -\sqrt{\omega}\, x \ , \hspace{1cm} p_2 \,=\,
-\frac{p_x}{\sqrt{\omega}} \ .
\end{eqnarray}
Assuming the symmetric Weyl ordering one obtains the following representation of  the quantum
Bateman's Hamiltonian (\ref{HHH}) with
\begin{equation}\label{}
\hat{H}_0 = \omega( \hat{p}_1\hat{x}_2 - \hat{p}_2\hat{x}_1)\ ,
\end{equation}
and
\begin{equation}\label{}
    \hat{H}_I = - \frac\gamma 2(\hat{p}_1\hat{x}_1
  + \hat{x}_1\hat{p}_1 + \hat{p}_2\hat{x}_2 + \hat{x}_2\hat{p}_2)\
  .
\end{equation}

\section{Spectral properties of the Hamiltonian}
\label{New-Rep} \setcounter{equation}{0}

\subsection{Polar representation}

The formula (\ref{H-osc-d}) for $H$ considerably simplifies in
polar coordinates:
\begin{equation*}\label{}
  x_1 + ix_2 = r e^{i\varphi}\ .
\end{equation*}
Defining the corresponding conjugate momenta
\begin{equation}\label{}
    p_\varphi = L_3\ , \hspace{1cm} p_r = \frac{{\bf x}{\bf
    p}}{r}\ ,
\end{equation}
with $L_3$ denoting 3rd component of ${\bf L} = {\bf x} \times
{\bf p}$ in $ \mathbb{R}^3$, one finds
\begin{equation}\label{H-radial-class}
H = -\omega p_\varphi - \gamma rp_r\ .
\end{equation}
The Hamilton equations in polar representation have the following
simple form:
\begin{equation}\label{}
    \dot{\varphi} = -\omega\ , \hspace{1cm} \dot{p}_\varphi=0\ ,
\end{equation}
and
\begin{equation}\label{}
    \dot{r} = - \gamma r \ , \hspace{1cm} \dot{p}_r = \gamma p_r\
    .
\end{equation}
The polar representation nicely shows that the Hamiltonian
dynamics consists in pure oscillation in $\varphi$--sector and
dissipation (pumping) in $r$--sector ($p$--sector). In our opinion
it is the most convenient representation to deal with .

The quantization of (\ref{H-radial-class}) leads to (\ref{HHH}) with
\begin{equation}\label{H-osc-d0}
\hat{H}_0 = - \omega \hat{p}_\varphi = i \omega\hbar \, \frac{\partial}{\partial \varphi}\ ,
\end{equation}
and
\begin{equation}\label{}
 \hat{H}_I = i \gamma\hbar  \left(
 r\frac{\partial}{\partial r} + 1 \right ) = -\gamma \left( r\hat{p}_r - \frac{i\hbar}{2} \right)  \ ,
\end{equation}
where the radial momentum $\hat{p}_r$ is defined by
\begin{equation}\label{pr}
\hat{p}_r = - i\hbar \left( \frac{\partial}{\partial r} + \frac{1}{2r} \right) \ .
\end{equation}
One easily finds the polar representation of the $su(1,1)$ generators:
\begin{eqnarray}
\hat{J}_1 &=& -\frac{\hbar}{4}\Big(\frac{\partial^2}{\partial
r^2}+\frac{1}{r^2}\frac{\partial^2}{\partial
\phi^2}\Big)-\frac{1}{4\hbar}r^2\,,\\
\hat{J}_2 &=& \frac i2 \Big(r\frac{\partial}{\partial r}+1\Big)\,,\\
\hat{J}_3 &=& \frac{1}{4\hbar}r^2+\frac i2
\frac{\partial}{\partial\phi}-\frac{\hbar}{4}
\Big(\frac{\partial^2}{\partial
r^2}+\frac{1}{r^2}\frac{\partial^2}{\partial \phi^2}\Big)\,.
\end{eqnarray}
together with the Casimir operator
\begin{equation}\label{}
\hat{J}_0 = \frac i2  \frac{\partial}{\partial\phi}\ .
\end{equation}
Note, that unitary evolution generated by $\hat{H}$ is given by
\begin{equation}\label{}
    \hat{U}(t) = e^{-i\hat{H}t/\hbar} = e^{-i\hat{H}_0 t/\hbar} \, e^{-i\hat{H}_I t/\hbar} =
    e^{\gamma t}\, 
\exp\left(\omega t  \frac{\partial}{\partial \varphi}\right)
    \exp\left(\gamma t\, r \frac{\partial}{\partial r}\right) \ ,
\end{equation}
and hence
\begin{equation}\label{U-t}
 (\hat{U}(t)\psi)(r,\varphi) = e^{\gamma t} \psi( e^{\gamma t}r, \varphi + \omega t) \ .
\end{equation}

\subsection{Complete set of eigenvectors}

It is evident that $\hat{H}$ defines an unbounded operator in
${\cal H} = L^2( \mathbb{R}^2,dx_1dx_2)$. It has continuous
spectrum $\sigma(\hat{H}) = (-\infty,\infty)$. To find the
corresponding generalized eigenvectors let us note that in polar
representation the Hilbert space $\cal H$ of square integrable
functions in $ \mathbb{R}^2$ factorizes as follows:
\begin{equation}\label{}
L^2( \mathbb{R}^2,dx_1dx_2) = L^2([0,2\pi), d\varphi) \otimes L^2(
\mathbb{R}_+,rdr) \ .
\end{equation}
Therefore, the spectral problem splits into two separate problems
in $L^2([0,2\pi),d\varphi)$ and $L^2( \mathbb{R}_+,rdr)$. One
easily finds
\begin{equation}\label{}
    \hat{H}\Psi_{l\lambda} = E_{l\lambda}\Psi_{l\lambda}\ ,
\end{equation}
with
\begin{equation}\label{E-n-lambda}
    E_{l\lambda} = \hbar(l\omega + \lambda\gamma)\,.
\end{equation}
The corresponding eigenvectors $\Psi_{l\lambda}$ are defined by
\begin{equation}\label{Psi-n-lambda}
    \Psi_{l\lambda}(r,\varphi) = \Phi_l(\varphi)\,
    R_\lambda(r)\ ,
\end{equation}
where
\begin{equation}\label{Phi-n}
  \Phi_l(\varphi) := \frac{e^{-i l
  \varphi}}{\sqrt{2\pi}}\ , \qquad l =0,\pm 1,\pm 2,\ldots \, ,
\end{equation}
and
\begin{equation}\label{R-lambda}
    R_\lambda(r) =  \frac{r^{-(i\lambda+1)}}{\sqrt{2\pi}} \ , \
    \ \ \ \ \lambda \in \mathbb{R}\ .
\end{equation}
Note, that $\Phi_l \in L^2([0,2\pi), d\varphi)$ whereas
$R_\lambda$ does not belong to $L^2( \mathbb{R}_+,rdr)$.

One easily shows that  the family $\Psi_{l\lambda}$ satisfies
\begin{equation}
    \int_0^{2\pi}\int_0^\infty \overline{\Psi_{l\lambda}}(r,\varphi)\, \Psi_{l'\lambda'}
    (r,\varphi)\, r\,dr\,d\varphi =
    \delta_{ll'}\delta(\lambda-\lambda')\, ,
\end{equation}
and
\begin{equation}\label{}
   \sum_{l=-\infty}^\infty \int_{-\infty}^\infty
   \overline{\Psi_{l\lambda}}(r,\varphi)\, \Psi_{l\lambda}(r',\varphi')\, d\lambda =
    \frac 1r\, \delta(r-r')\delta(\varphi-\varphi')\ .
\end{equation}
They imply the following resolution of identity
\begin{equation}\label{I><}
    \oper = \sum_{l=-\infty}^\infty \int_{-\infty}^\infty\,
    d\lambda\, |\Psi_{l\lambda}\rangle \langle \Psi_{l\lambda}| \,,
\end{equation}
and the spectral resolution of  Hamiltonian
\begin{equation}\label{H><}
\hat{H} = \sum_{l=-\infty}^\infty \int_{-\infty}^\infty\,
    d\lambda\, E_{l\lambda} |\Psi_{l\lambda}\rangle \langle \Psi_{l\lambda}|\,,
\end{equation}

\subsection{Feynman propagator}

Let us calculate the corresponding Feynman propagator
\begin{equation}
    K({\bf x},t|{\bf x}',t') = \la {\bf x}|\hat{U}(t-t')|{\bf x}'\r\ ,
\end{equation}
where $\hat{U}(\tau) = \exp(-i\hat{H}\tau/\hbar)$. Using polar representation one finds
\begin{equation}
    K(r,\varphi,t|r',\varphi',t') = \sum_{l=-\infty}^\infty \int_{-\infty}^\infty
    e^{-iE_{l\lambda}\tau/\hbar}\,
    \Psi_{l\lambda}(r,\varphi)\,\overline{\Psi_{l\lambda}}(r',\varphi')\, d\lambda\ ,
\end{equation}
with $\tau = t - t'$.
Now, using (\ref{Psi-n-lambda}) one obtains
\begin{equation}\label{}
K(r,\varphi,t|r',\varphi',t') = K_1(r,t|r',t')K_2(\varphi,t|\varphi',t')\ ,
\end{equation}
where the radial  and azimuthal propagators are given by
\begin{equation}\label{}
K_1(r,t|r',t') = \int_{-\infty}^\infty e^{-i\lambda\gamma\tau}\,R_\lambda(r)
\overline{R_\lambda}(r')\, d\lambda\ ,
\end{equation}
and
\begin{equation}\label{}
K_2(\varphi,t|\varphi',t') = \sum_{l=-\infty}^\infty e^{-i\omega
l\tau}
 \Phi_l(\varphi)\, \overline{\Phi_l}(\varphi')\,,
\end{equation}
respectively.
Finally, formulae (\ref{Phi-n}) and (\ref{R-lambda}) imply
\begin{equation}\label{}
K_2(\varphi,t|\varphi',t') = \delta(\varphi' - \varphi - \omega\tau)\ ,
\end{equation}
and
\begin{eqnarray}\label{}
K_1(r,t|r',t') &=& \frac{1}{2\pi} \frac{1}{rr'}\, \int_{-\infty}^\infty\, e^{i\lambda(\ln r' - \ln r - \gamma\tau)}\,
d\lambda\ \nonumber \\
&=& \frac{1}{rr'}\, \delta(\ln r' - \ln r - \gamma\tau) = e^{\gamma\tau}\, \frac{
\delta(r' - re^{\gamma\tau})}{r'}\ .
\end{eqnarray}
Therefore, the time evolution is given by
\begin{eqnarray}\label{}
    \psi_t(r,\varphi) &=& \int_0^{2\pi}\int_0^\infty \,
    K(r,\varphi,t|r',\varphi',t'=0)\, \psi_0(r',\varphi')\,
    r'dr'd\varphi' \nonumber \\
    &=& e^{\gamma t} \, \psi_0(e^{\gamma t}r, \varphi + \omega t)\ ,
\end{eqnarray}
which perfectly agrees with (\ref{U-t}).

\section{Analyticity and complex eigenvalues} \label{Analyticity}
\setcounter{equation}{0}

Now we are going to relate the energy eigenvectors
$\Psi_{n\lambda}$ corresponding to the real spectrum
$E_{n\lambda}$ with the family of discrete complex eigenvalues of
the Bateman's Hamiltonian. Let us consider the distribution
$\Psi_{n\lambda}$ with $\lambda \in \mathbb{C}$, i.e. for any test
function $\phi(r,\varphi)$
\begin{equation}\label{}
    \Psi_{l\lambda}(\phi) = \langle \phi|\Psi_{l\lambda} \rangle =
    \int_0^\infty r^{-i\lambda} \overline{\phi_l}(r)\, dr\,,
\end{equation}
where
\begin{equation}\label{phi-n}
    \phi_l(r) = \frac{1}{2\pi} \int_0^{2\pi}
    e^{il\varphi}\phi(r,\varphi)\, d\varphi\, .
\end{equation}
Now, the analytical properties of $\Psi_{l\lambda}$ depend upon
the behavior  of $\phi_l(r)$ at $r=0$. A distribution $r^\alpha$
acting on the space of smooth functions $S( \mathbb{R}_+)$
\begin{equation}\label{}
S( \mathbb{R}_+) \ni f \ \longrightarrow\ \int_0^\infty r^\alpha
\overline{f}(r)\ dr \ ,
\end{equation}
is well defined for all $\alpha \in \mathbb{C}$ except the
discrete family of points  where it may have simple poles (see
e.g.~\cite{Gel-S}). The location of poles depends upon the
behavior of a test function $f$ at $r=0$. Assuming the most
general expansion of $f(r)$
\begin{equation}\label{}
    f(r) = f_0 + f_1\, r + f_2\,  r^2 + \ldots \ ,
\end{equation}
the poles are located at $\alpha = -1,-2,-3, \ldots .$ However,
$\phi_l(r)$ defined in (\ref{phi-n}) is much more regular. It can
be observed (see Appendix B.) that $\phi_l(r)$ may be expanded at
$r=0$ as follows:
\begin{equation}\label{phi-n_expansion}
    \phi_l(r) = a_l \,r^{|l|} + a_{l+2}\, r^{|l|+2} + a_{l+4}\,
    r^{|l|+4} + \ldots \ .
\end{equation}
Therefore, the poles that remain are located at
\begin{equation}\label{lambda-nl}
    \lambda_{nl} = - i(|l| + 2n +1) \ , \hspace{1cm}
    n=0,1,2,\ldots\ .
\end{equation}
Moreover, the corresponding residues of $\Psi_{l\lambda}$ are
given by
\begin{equation}\label{Res1}
    {\rm  Res}\, \Psi_{l\lambda}\Big|_{\lambda=\lambda_{nl}} =
     \frac{1}{\sqrt{(|l|+2n)!}} \  \frac{\mathfrak{f}^-_{nl}}{\sqrt{2\pi}}\ ,
\end{equation}
where
\begin{equation}\label{f-nl}
 \mathfrak{f}^-_{nl}(r,\varphi) =  \Phi_l(\varphi)\,\frac{i(-1)^{|l|+2n}}{\sqrt{(|l|+2n)!}}\,
  \frac{ \delta^{(|l|+2n)}(r)}{r}\ .
\end{equation}
On the other hand
\begin{equation}\label{Res11}
     \overline{\Psi_{l\lambda}}\Big|_{\lambda=\lambda_{nl}} =
    \sqrt{(|l|+2n)!} \,  \frac{\overline{\mathfrak{f}^+_{nl}}}{\sqrt{2\pi}}\ ,
\end{equation}
where
\begin{equation}\label{f+nl}
 \mathfrak{f}^+_{nl}(r,\varphi) =  \Phi_l(\varphi)\,\frac{r^{|l|+2n}}{\sqrt{(|l|+2n)!}} \,.
\end{equation}
Now, the crucial observation is that $\mathfrak{f}^\pm_{nl}$  satisfy
\begin{equation}\label{J0-f}
    \hat{J}_0\, |\mathfrak{f}^\pm_{nl}\rangle = \frac l2\, |\mathfrak{f}^\pm_{nl}\rangle\ ,
\end{equation}
and
\begin{equation}\label{J2-f}
\hat{J}_2\, |\mathfrak{f}^\pm_{nl}\rangle = \pm \frac i2 (|l| + 2n
+1)\,|\mathfrak{f}^\pm_{nl}\rangle\, ,
\end{equation}
which proves that they define eigenvectors of $\hat{H}$
\begin{equation}\label{}
    \hat{H}|\mathfrak{f}^\pm_{nl}\rangle = E^\pm_{nl} |\mathfrak{f}^\pm_{nl}\rangle \, ,
\end{equation}
corresponding to complex eigenvalues
\begin{equation}\label{Enl}
    E^\pm_{nl} = \hbar\omega l \pm i \hbar \gamma (|l| + 2n +1)\, .
\end{equation}
The above formula is equivalent to the Bateman's spectrum (\ref{FTspectrum}) after
the following identification
\begin{equation}
    j = \frac l 2 \, ,
\end{equation}
and
\begin{equation}\label{}
    m = \frac 12 ( |l| + 2n) = |j| + n \, ,
\end{equation}
which reproduces condition (\ref{jm!!!}). In terms of $(n_A,n_B)$ one has
\begin{eqnarray}\label{}
n_A &=& \frac 12 \, ( |l| + l) + n \, , \\
n_B &=&  \frac 12 \,( |l| - l) + n \, .
\end{eqnarray}
We have therefore the following relation between $|\psi_{jm}^\pm\r$ and
$|\mathfrak{f}^\pm_{nl}\rangle$:
\begin{equation}
    |\psi^\pm_{jm}\r = |\mf^\pm_{2j,m-|j|}\r\ ,
\end{equation}
that is, $|\mathfrak{f}^\pm_{nl}\rangle$ defined in (\ref{f-nl}) and (\ref{f+nl}) may be
regarded as a particular representation of $|\psi_{jm}^\pm\r$.

Let us introduce two important classes of functions  \cite{Duren}: consider
the space of complex functions $f : \mathbb{C}\rightarrow  \mathbb{C}$.
A smooth function $f=f(\lambda)$ is in the  Hardy class from above
${\cal H}^2_+$ (from below ${\cal H}^2_-$) if $f(\lambda)$ is a boundary
value of an analytic function  in the upper, i.e. $\mbox{Im}\,
\lambda\geq 0$ (lower, i.e. $\mbox{Im}\, \lambda\leq 0$) half complex
$\lambda$-plane vanishing faster than any power of $\lambda$ at the upper
(lower) semi-circle $|\lambda| \rightarrow \infty$. Now, define
\begin{eqnarray}  \label{S-}
{\cal S}_- = \Big\{ \phi \in {\cal S}\, \Big| \, \la
\Psi_{l\lambda} | \phi \r \in {\cal H}^2_-\, \Big\} \ ,
\end{eqnarray}
that is, $\phi \in {\cal S}_-$ iff the complex function
\[  \mathbb{C}\ni \lambda \ \longrightarrow \ \la
\Psi_{l\lambda} | \phi \r \, \in \mathbb{C}\ , \] is in the Hardy
class from below $ {\cal H}^2_-$. Equipped with this mathematical
notion let us consider an arbitrary test function $\phi \in {\cal
S}_-$. The resolution of identity (\ref{I><}) implies
\begin{eqnarray}
\phi(r,\varphi) = \sum_{l=-\infty}^\infty \int_{-\infty}^\infty
d\lambda\, \Psi_{l\lambda}(r,\varphi) \la \Psi_{l\lambda} | \phi\r
\, .
\end{eqnarray}
Now, since $ \la \Psi_{l\lambda} | \phi\r \in {\cal H}^2_-$, we
may close the integration contour along the lower semi-circle
$|\lambda|\rightarrow \infty$ (see Figure~1).

\begin{figure}[t]
\begin{center}
\epsfig{figure=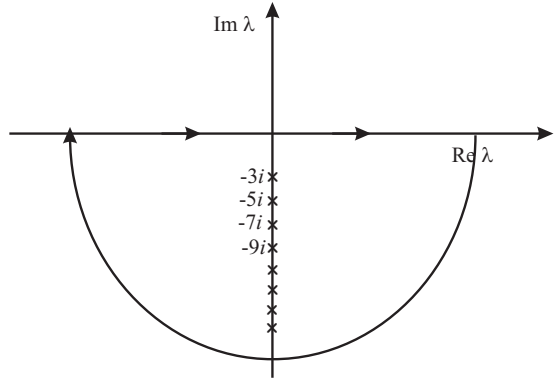,width=0.5\textwidth}
\end{center}
\caption{Integration contour along the lower semi-circle for
$l=2$.}
\end{figure}

Hence, due to the residue theorem one
obtains
\begin{eqnarray} \label{phi-R}
\phi(r,\varphi) = -2\pi i \sum_{l=-\infty}^\infty
\sum_{n=0}^\infty \mbox{Res}\,
\Psi_{l\lambda}(r,\varphi)\Big|_{\lambda=\lambda_{nl}}\, \la
\Psi_{l\lambda} | \phi\r\Big|_{\lambda=\lambda_{nl}} \, .
\end{eqnarray}
Finally, using (\ref{Res1}) and (\ref{Res11}) one gets
\begin{equation}
\phi(r,\varphi) = \sum_{l=-\infty}^\infty \sum_{n=0}^\infty\,
     \mf^-_{nl}(r,\varphi)\, \la \mf^+_{nl}|\phi\r \, .
\end{equation}
We have proved, therefore, that the subspace ${\cal S}_- \subset {\cal S} \subset {\cal H}$ gives
rise to the following resolution of identity
\begin{equation}\label{I-S-}
  \oper_- \equiv  \oper\Big|_{{\cal S}_-} \, = \, \sum_{l=-\infty}^\infty \sum_{n=0}^\infty\,
     |\mf^-_{nl}\r \la \mf^+_{nl}| \, .
\end{equation}
The same arguments lead us to the following spectral resolution of $\hat{H}$ restricted to ${\cal S}_-$:
\begin{equation}\label{H-S-}
    \hat{H}_- \equiv
\hat{H}\Big|_{{\cal S}_-} \, = \, \sum_{l=-\infty}^\infty
\sum_{n=0}^\infty\,
     E^-_{nl}\,|\mf^-_{nl}\r \la \mf^+_{nl}| \, ,
\end{equation}
with $E^-_{nl}$ defined in (\ref{Enl}). Introducing the following family of operators
\begin{equation}\label{P-nl-}
    \hat{P}^-_{nl} = |\mf^-_{nl}\r \la \mf^+_{nl}|\ ,
\end{equation}
the spectral decompositions (\ref{I-S-}) and (\ref{H-S-}) may be rewritten as follows
\begin{equation}
\oper_- = \sum_{l=-\infty}^\infty \sum_{n=0}^\infty\,
\hat{P}^-_{nl}\, ,
\end{equation}
and
\begin{equation}
\hat{H}_- = \sum_{l=-\infty}^\infty \sum_{n=0}^\infty\,
     E^-_{nl}\, \hat{P}^-_{nl}\, .
\end{equation}
Note, that
\begin{equation}\label{PPP-}
\hat{P}^-_{nl}\, \hat{P}^-_{n'l'}  = \delta_{nl} \delta_{n'l'}\,
\hat{P}^-_{nl} \,,
\end{equation}
that is, the family $\hat{P}^-_{nl}$ seems to play the role of the family of orthogonal projectors.
Note, however, that $\hat{P}^-_{nl}$ are not hermitian.

\section{Time reversal}
\label{Reversal}

It was shown in \cite{Bla04} that Bateman's Hamiltonian is time
reversal invariant
\begin{equation}\label{}
    {\cal T}^\dag \hat{H} {\cal T} = \hat{H}\ ,
\end{equation}
where $\cal T$ denote the anti-unitary time reversal operator.
Moreover, it turns out \cite{Bla04} that both $\hat{J}_0$ and
$\hat{J}_2$ satisfy
\begin{equation}\label{TJTJ}
    {\cal T}^\dag \hat{J}_0 {\cal T} = \hat{J}_0\ , \hspace{1cm}
    {\cal T}^\dag \hat{J}_2 {\cal T} = \hat{J}_2\ .
\end{equation}
Let us define
\begin{equation}\label{}
    \Xi_{l\lambda} = {\cal T}\Psi_{l\lambda}\,.
\end{equation}
In analogy with (\ref{I><}) and (\ref{H><}) one has the  following
 resolution of identity
\begin{equation}\label{I><T}
\oper = \sum_{l=-\infty}^\infty \int_{-\infty}^\infty\,
    d\lambda\, |\Xi_{l\lambda}\rangle \langle \Xi_{l\lambda}| \,,
\end{equation}
and spectral resolution of the Hamiltonian
\begin{equation}\label{H><T}
\hat{H} = \sum_{l=-\infty}^\infty \int_{-\infty}^\infty\,
    d\lambda\, E_{l\lambda}\, |\Xi_{l\lambda}\rangle \langle \Xi_{l\lambda}|
    \,.
\end{equation}
Now, let us introduce another subspace ${\cal S}_+$ in the space of test functions
\begin{eqnarray}  \label{S+}
{\cal S}_+ = \Big\{ \phi \in {\cal S}\, \Big| \, \la
\Xi_{l\lambda} | \phi \r \in {\cal H}^2_+\, \Big\} \, ,
\end{eqnarray}
that is, $\phi \in {\cal S}_+$ iff the complex function
\[  \mathbb{C}\ni \lambda \ \longrightarrow \ \la
\Xi_{l\lambda} | \phi \r \, \in \mathbb{C}\ , \] is in the Hardy
class from above $ {\cal H}^2_+$. It is easy to show that
\begin{equation}\label{}
    {\cal S}_+ = {\cal T}({\cal S}_-)\ ,
\end{equation}
and vice versa
\begin{equation}\label{}
    {\cal S}_- = {\cal T}({\cal S}_+)\ .
\end{equation}
Indeed, if $\phi \in {\cal S}_-$ then $\la \Psi_{l\lambda} | \phi
\r \in {\cal H}^2_-$. One has therefore
\begin{equation}
\la \Xi_{l\lambda} | {\cal T}\phi \r = \la \phi | {\cal T}^\dag\,
\Xi_{l\lambda} \r = \overline{\la \Psi_{l\lambda}|\phi\r} \in
{\cal H}^2_+\,,
\end{equation}
which implies that ${\cal T}\phi \in {\cal S}_+$.\footnote{In the above formulae we have
used
\[   \la \psi| {\cal A} \phi\r = \la \phi | {\cal A}^\dag \psi\r\,,  \]
which holds for any anti-linear operator $\cal A$.} Moreover
\begin{equation}\label{SS}
{\cal S}_- \cap {\cal S}_+ = \{ \emptyset \}\, .
\end{equation}
To prove this property let us assume that $\phi \in {\cal S}_-
\cap {\cal S}_+$. Since $\phi \in {\cal S}_+$, one has ${\la
\Xi_{l\lambda}|\phi\r}\in {\cal H}^2_+$. However
\begin{equation}
\la \Xi_{l\lambda}|\phi\r = \overline{\la \phi| {\cal
T}\Psi_{l\lambda}\r } = \la \Psi_{l\lambda}| {\cal T}^\dag \phi\r
\in {\cal H}^2_+\, .
\end{equation}
On the other hand ${\cal T}^\dag \in {\cal S}_-$ and hence $ \la
\Psi_{l\lambda}| {\cal T}^\dag \phi\r \in {\cal H}^2_-$.
Therefore, $\la \Psi_{l\lambda}| {\cal T}^\dag \phi\r \in {\cal
H}^2_- \cap {\cal H}^2_+$ which means that $\la \Psi_{l\lambda}|
{\cal T}^\dag \phi\r$ is en entire function vanishing on the
circle $|\lambda| \longrightarrow \infty$. However, any entire
function is necessarily bounded and hence such $\phi$ which
belongs both to ${\cal S}_-$ and ${\cal S}_+$ does not exist.

Now, take any test function $\phi \in {\cal S}_+$. Formula (\ref{I><T}) implies
\begin{eqnarray} \label{phi-int}
\phi(r,\varphi) = \sum_{l=-\infty}^\infty \int_{-\infty}^\infty
d\lambda\, \Xi_{l\lambda}(r,\varphi) \la \Xi_{l\lambda} | \phi\r
\,.
\end{eqnarray}
Let us continue the eigenvectors $\Xi_{l\lambda}$ for the complex
$\lambda$ plane. They have simple poles at $\lambda =
-\lambda_{nl}$ with $\lambda_{nl}$ defined in (\ref{lambda-nl}).
The corresponding residues of $\Xi_{l\lambda}$ follows from
(\ref{Res1})
\begin{equation}\label{Res21}
    {\rm  Res}\ \Xi_{l\lambda}\Big|_{\lambda=-\lambda_{nl}} =
     \frac{1}{\sqrt{(|l|+2n)!}} \  \frac{{\cal T}\,\mathfrak{f}^-_{nl}}{\sqrt{2\pi}}\ .
\end{equation}
Moreover,
\begin{equation}\label{Res22}
     \overline{\Xi_{l\lambda}}\Big|_{\lambda=-\lambda_{nl}} =
    \sqrt{(|l|+2n)!} \  \frac{{\cal T}\overline{\mathfrak{f}^+_{nl}}}{\sqrt{2\pi}}\, .
\end{equation}
Now, since $ \la \Xi_{n\lambda} | \phi\r \in {\cal H}^2_+$, we may
close the integration contour in (\ref{phi-int}) along the upper
semi-circle $|\lambda|\rightarrow \infty$. The residue theorem
implies
\begin{eqnarray} \label{phi-Res}
\phi(r,\varphi) = 2\pi i \sum_{l=-\infty}^\infty \sum_{n=0}^\infty
\mbox{Res}\
\Xi_{l\lambda}(r,\varphi)\Big|_{\lambda=-\lambda_{nl}}\, \la
\Xi_{l\lambda} | \phi\r\Big|_{\lambda=-\lambda_{nl}} \, .
\end{eqnarray}
Finally, using (\ref{Res21}) and (\ref{Res22}) one gets
\begin{equation}\label{}
\phi(r,\varphi) = \sum_{l=-\infty}^\infty \sum_{n=0}^\infty\,
     {\cal T}\mf^-_{nl}(r,\varphi)\, \overline{\la \phi|{\cal T}\mf^+_{nl}}\r \,,
\end{equation}
and hence it implies the following resolution of identity on ${\cal S}_+$:
\begin{equation}\label{I-S+}
    \oper\Big|_{{\cal S}_+} \, = \, \sum_{l=-\infty}^\infty \sum_{n=0}^\infty\,
     {\cal T}|\mf^-_{nl}\r \la \mf^+_{nl}|{\cal T}^\dag \, .
\end{equation}
Now, the formula (\ref{J2-f}) together with (\ref{TJTJ}) gives
\begin{equation}\label{J2-Tf}
\hat{J}_2\, {\cal T}|\mathfrak{f}^\pm_{nl}\rangle = \mp
\frac{i}2(|l| + 2n +1)\,{\cal T} |\mathfrak{f}^\pm_{nl}\rangle\ ,
\end{equation}
and hence one deduces the following relations between $|\mathfrak{f}^\pm_{nl}\rangle$ and time reversed
${\cal T}|\mathfrak{f}^\pm_{nl}\rangle$
\begin{equation}\label{T-f-alpha}
    {\cal T}|\mathfrak{f}^+_{nl}\rangle = e^{i\alpha_{nl}}\,|\mathfrak{f}^-_{nl}\rangle\, ,
    \hspace{1cm}
 {\cal T}|\mathfrak{f}^-_{nl}\rangle = e^{i\alpha_{nl}}\,|\mathfrak{f}^+_{nl}\rangle\, ,
\end{equation}
where $\alpha_{nl}$ are arbitrary $(n,l)$-depended phases. It should be stressed that these phases are
physically irrelevant. Actually, one may redefine $|\mathfrak{f}^+_{nl}\rangle$ in (\ref{f-nl}) and
(\ref{f+nl}) such that these additional phase factors disappear from (\ref{T-f-alpha}). Let us observe
that
\begin{equation}\label{}
{\cal T}^2|\mathfrak{f}^\pm_{nl}\rangle = |\mathfrak{f}^\pm_{nl}\rangle\ ,
\end{equation}
irrespective of $\alpha_{nl}$. Taking into account (\ref{T-f-alpha}) one obtains from (\ref{I-S+})
\begin{equation}\label{I-S+final}
   \oper_+ \equiv \oper\Big|_{{\cal S}_+} \, = \, \sum_{l=-\infty}^\infty \sum_{n=0}^\infty\,
     |\mf^-_{nl}\r \la \mf^+_{nl}|\, .
\end{equation}
The same arguments lead us to the following spectral resolution of $\hat{H}$
\begin{equation}\label{H-S+final}
\hat{H}_+ \equiv    \hat{H}\Big|_{{\cal S}_+} \, = \,
\sum_{l=-\infty}^\infty \sum_{n=0}^\infty\,
     E^+_{nl}\,|\mf^+_{nl}\r \la \mf^-_{nl}| \, ,
\end{equation}
with $E^+_{nl}$ defined in (\ref{Enl}). Finally, introducing
\begin{equation}\label{P-nl+}
    \hat{P}^+_{nl} = |\mf^+_{nl}\r \la \mf^-_{nl}| = ( \hat{P}^-_{nl})^\dag\ ,
\end{equation}
with $ \hat{P}^-_{nl}$ defined in (\ref{P-nl-}),
the spectral decompositions (\ref{I-S+final}) and (\ref{H-S+final}) may be rewritten as follows
\begin{equation}\label{}
\oper_+ = \sum_{l=-\infty}^\infty \sum_{n=0}^\infty\,
\hat{P}^+_{nl}\,,
\end{equation}
and
\begin{equation}\label{}
\hat{H}_+ = \sum_{l=-\infty}^\infty \sum_{n=0}^\infty\,
     E^+_{nl}\, \hat{P}^+_{nl}\, .
\end{equation}

\section{Resonances  and dissipation}

What is the physical meaning of the complex eigenvalues
$E^\pm_{nl}$? To answer this question let us consider the
resolvent operator of the Bateman's Hamiltonian
\begin{equation}
    \hat{\rm R}(\hat{H},z) = (\hat{H} - z)^{-1}\, .
\end{equation}
Using the family of eigenfunctions $|\Psi_{l\lambda}\r$ one has
\begin{equation}\label{}
 \hat{\rm R}(\hat{H},z) = \sum_{l=-\infty}^\infty \int_{-\infty}^\infty\,
    \, \frac{d\lambda}{E_{l\lambda}-z}\ |\Psi_{l\lambda}\rangle \langle \Psi_{l\lambda}|
    \,,
\end{equation}
with $E_{l\lambda}$ defined in (\ref{E-n-lambda}). Now, using the
same technique as in Section~\ref{Analyticity} one easily finds
\begin{equation}\label{}
    \hat{\rm R}_-(z) \equiv  \hat{\rm R}(\hat{H},z)\Big|_{S_-} = \sum_{l=-\infty}^\infty
    \sum_{n=0}^\infty\,
     \frac{1}{E^-_{nl}-z}\, \hat{P}^-_{nl}\, ,
\end{equation}
with $P^-_{nl}$ defined in (\ref{P-nl-}). This shows that
$E^-_{nl}$ constitute poles of the resolvent operator on ${\cal
S}_-$. In the same way using the family $|\Xi_{l\lambda}\r$
\begin{equation}\label{}
 \hat{\rm R}(\hat{H},z) = \sum_{l=-\infty}^\infty \int_{-\infty}^\infty\,
    \, \frac{d\lambda}{E_{l\lambda}-z}\ |\Xi_{l\lambda}\rangle \langle \Xi_{l\lambda}|
    \,,
\end{equation}
one finds
\begin{equation}\label{}
   \hat{\rm R}_+(z) \equiv    \hat{\rm R}(\hat{H},z)\Big|_{S_+} =
   \sum_{l=-\infty}^\infty \sum_{n=0}^\infty\,
     \frac{1}{E^+_{nl}-z}\ \hat{P}^+_{nl}\, ,
\end{equation}
which shows that $E^+_{nl}$ constitute poles of the resolvent
operator on ${\cal S}_+$. As is well known \cite{RES-1} the poles
of the resolvent operator correspond to resonant states. Hence,
the complex eigenvalues $E^\pm_{nl}$  may be interpreted as
resonances of the Bateman's Hamiltonian. Note that due to the
Cauchy theorem operators $ \hat{P}^\pm_{nl} $ may be represented
by the following integrals
\begin{equation}\label{}
    \hat{P}^\pm_{nl} =  \frac{1}{2\pi i}\, \oint_{\gamma^\pm_{nl}} \hat{\rm R}_\pm(z)\, dz\, ,
\end{equation}
where $\gamma^\pm_{nl}$ is any  (clockwise) closed curve which
encircles a single pole $z = E^\pm_{nl}$ (see Figure~2).

\begin{figure}[t]
\begin{center}
\epsfig{figure=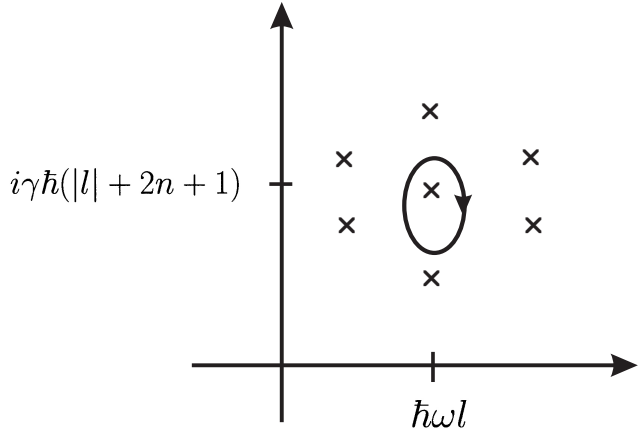,width=0.6\textwidth}
\end{center}
\caption{A closed curve $\gamma_{nl}^+$ on a complex energy
plane.}
\end{figure}

Finally, let us turn to the evolution generated by the Bateman's
Hamiltonian. Clearly,
\[ \mathbb{R} \ni t \ \longrightarrow\ \hat{U}(t) = \exp(-i\hat{H}t/\hbar)\ , \]
defines a group of unitary operators on the Hilbert space $ L^2( \mathbb{R}^2)$.
Now, it is easy to see that if $\psi_- \in {\cal S}_-$, then $\hat{U}(t)\psi_-$ belongs to ${\cal S}_-$
only if $t \geq 0$. Similarly, if $\psi_+ \in {\cal S}_+$, then $\hat{U}(t)\psi_+$ belongs to ${\cal S}_+$
only if $t \leq 0$. Therefore, we have two natural semigroups
\begin{equation}
    \hat{U}_-(t)\ :\ {\cal S}_-\ \longrightarrow\ {\cal S}_-\ , \hspace{1cm} {\rm for} \ \ t\geq 0 \ ,
\end{equation}
and
\begin{equation}
    \hat{U}_+(t)\ :\ {\cal S}_+\ \longrightarrow\ {\cal S}_+\ , \hspace{1cm} {\rm for} \ \ t\leq 0 \ ,
\end{equation}
where
\begin{equation}
 \hat{U}_-(t) = \hat{U}(t)\Big|_{{\cal S}_-} \ , \hspace{.7cm} {\rm and}  \hspace{.7cm}
\hat{U}_+(t) = \hat{U}(t)\Big|_{{\cal S}_+} \ .
\end{equation}
One has
\begin{equation}
    \psi_-(t) =  \hat{U}_-(t)\psi_- =  \sum_{l=-\infty}^\infty e^{-i\omega lt}\,
    \sum_{n=0}^\infty\,
    e^{-\gamma(|l|+n+1)t}\, \hat{P}^-_{nl}\,\psi_-\, ,
\end{equation}
for $t\geq 0$, and
\begin{equation}
    \psi_+(t) =  \hat{U}_+(t)\psi_+ =  \sum_{l=-\infty}^\infty e^{-i\omega lt}\,
    \sum_{n=0}^\infty\,
    e^{\gamma(|l|+n+1)t}\, \hat{P}^+_{nl}\,\psi_+\, ,
\end{equation}
for $t\leq 0$. It should be clear that these two semigroups are related by the time reversal operator $\cal T$:
indeed formulae (\ref{T-f-alpha}) imply
\begin{equation}
    {\cal T}\, \hat{P}^-_{nl} \, {\cal T}^\dag = \hat{P}^+_{nl}\
    \hspace{.7cm} {\rm and}  \hspace{.7cm}
    {\cal T}\, \hat{P}^+_{nl} \, {\cal T}^\dag = \hat{P}^-_{nl}\ ,
\end{equation}
and hence
\begin{eqnarray}
{\cal T}\, \hat{U}_-(t) \, {\cal T}^\dag &=&
\sum_{l=-\infty}^\infty \sum_{n=0}^\infty\,
{\cal T}\,\left( e^{-iE^-_{nl}t/\hbar}\, \hat{P}^-_{nl} \right)\, {\cal T}^\dag \nonumber\\
 &=& \sum_{l=-\infty}^\infty \sum_{n=0}^\infty\, e^{-iE^+_{nl}(-t)/\hbar}\, \hat{P}^+_{nl}
 = \hat{U}_+(-t)\ ,
\end{eqnarray}
for $t\geq 0$. Similarly, one finds
\begin{equation}\label{}
{\cal T}\, \hat{U}_+(t) \, {\cal T}^\dag = \hat{U}_-(-t)\ ,
\end{equation}
for $t\leq 0$. We have shown that perfectly reversible quantum dynamics $\hat{U}(t)$ on the full
Hilbert space $L^2( \mathbb{R}^2)$ is no longer reversible when restricted to the subspaces ${\cal S}_-$ and
${\cal S}_+$. This effective irreversibility is caused by the presence of resonant states $|\mf^\pm_{nl}\r$
corresponding to complex eigenvalues $E^\pm_{nl}$.

\section{Conclusions}
\label{CON}

 In this paper we have studied the spectral properties
of the Bateman Hamiltonian. It was shown that the complex
eigenvalues $E^\pm_{jm}$ given by (\ref{FTspectrum}) corresponds
to the poles of the resolvent operator $\hat{\rm R}(\hat{H},z) =
(\hat{H} - z)^{-1}$. Therefore, the corresponding generalized
eigenvectors may be interpreted as resonant states of the Bateman
dual system. It proves that dissipation and irreversibility is
caused by the presence of resonances.

 From the mathematical point of view
the Bateman system gives rise to the so called Gel'fand triplet or
rigged Hilbert space \cite{RHS1,RHS2} (see also
\cite{Bohm-Gadella,Bohm}). A Gel'fand triplet (rigged Hilbert
space) is a collection of spaces
\begin{equation}\label{}
    \Phi \subset {\cal H} \subset \Phi'\ ,
\end{equation}
 where $\cal H$ is a Hilbert space, $\Phi$  its dense subspace and $\Phi'$
 is the dual space of continuous linear functionals on $\Phi$.
 Note, that elements from $\Phi'$ do not belong to $\cal H$. This
 is a typical situation when one deals with the continuum
 spectrum. The corresponding generalized eigenvectors are no
 longer elements from the system Hilbert space. They are elements from the dual space
 $\Phi'$, i.e. distributions acting on $\Phi$ \cite{Schwartz,Yosida}. In  our case we
 have two natural Gel'fand triplets:
 \begin{equation}\label{}
    {\cal S}_- \subset L^2( \mathbb{R}^2) \subset {\cal S}_-'\ ,
\end{equation}
and
\begin{equation}\label{}
    {\cal S}_+ \subset L^2( \mathbb{R}^2) \subset {\cal S}_+'\ .
\end{equation}
The first triplet corresponds to the forward dynamics $\hat{U}_-$
and the second one  corresponds to the backward semigroup
$\hat{U}_+$. A similar analysis based on rigged Hilbert space
approach was performed in \cite{damp1,damp2} for a toy model
damped system defned by $\dot{x} = -\gamma x$.

\section*{Appendix A}
\def\theequation{A.\arabic{equation}}

Let us briefly sketch calculations leading to (\ref{Res1}) and
(\ref{Res11}). We introduce a distribution $\Psi_{l\lambda}$
acting on a test function $\phi(r,\varphi)$ as an antilinear
functional defined by the integral
\begin{equation}
\Psi_{l\lambda}(\phi)=\<\phi|\Psi_{l\lambda}\>=
\frac1{2\pi}\int_{\mathbb{R}^2}e^{-il\varphi}r^{-i\lambda-1}
\overline{\phi}(r,\varphi)dS=\int_0^\infty
r^{-i\lambda}\overline{\phi}_l(r)dr\,,
\end{equation}
where $\lambda\in\mathbb{C}\,$, $dS=rdr\,d\varphi$, and
$\phi_l(r)$ is given by (\ref{phi-n}). Expanding $\phi_l(r)$ in
the power series and rewriting the last integral as
\begin{eqnarray}
\int_0^\infty r^{-i\lambda}\overline{\phi}_l(r)dr &=& \int_0^1
r^{-i\lambda}\Big[\overline{\phi}_l(r)-
\overline{\phi}_l(0)-r\overline{\phi}_l'(0)-\ldots-\frac{r^{l-1}}{(l-1)!}
\overline{\phi}_l^{(l-1)}(0)\Big]dr\nonumber\\
&+&\int_1^\infty r^{-i\lambda}\overline{\phi}_l(r)dr \\
&+& \int_0^1
r^{-i\lambda}\Big[\overline{\phi}_l(0)+r\overline{\phi}_l'(0)+\ldots+
\frac{r^{l-1}}{(l-1)!}\overline{\phi}_l^{(l-1)}(0)\Big]dr \ ,
\nonumber
\end{eqnarray}
one can observe that the first two summands are regular for all
$\lambda\in\mathbb{C}$. The last integral, however, equals to
\begin{eqnarray}
\sum_{k=0}^{l-1}\frac{\overline{\phi}_l^{(k)}(0)}{k!}\int_0^1
r^{-i\lambda}r^k\,dr=
\sum_{k=0}^{l-1}\frac{\overline{\phi}_l^{(k)}(0)}{k!}\frac1{k-i\lambda+1}\label{expansion}
\end{eqnarray}
and has simple poles in $\lambda_k=-i(k+1)$, $k=0,1,\ldots,l-1$.
Moreover, one can read from (\ref{expansion}) that
\begin{equation}
{\rm Res}\,\<\phi|\Psi_{l\lambda}\>\Big|_{\lambda=-i(k+1)}=
\frac{\overline{\phi}_l^{(k)}(0)}{k!}.
\end{equation}
Finally, using
\begin{equation}
\overline{\phi}_l^{(k)}(0)=\frac1{2\pi}\int_0^{2\pi}
e^{-il\varphi}\overline{\phi}^{(k)}(0,\varphi)\,d\varphi
\end{equation}
and
\begin{equation}
\overline{\phi}^{(k)}(0,\varphi)=(-1)^k\int_0^\infty
\frac{\delta^{(k)}(r)}{r}\overline{\phi} (r,\varphi)\,rdr\ ,
\end{equation}
we get
\begin{equation}
\overline{\phi}_l^{(k)}(0)=\frac{(-1)^k}{2\pi}\int_{\mathbb{R}^2}e^{-il\varphi}
\frac{\delta^{(k)}(r)}{r}\overline{\phi} (r,\varphi)dS\ .
\end{equation}
But due to (\ref{phi-n_expansion}) in the case investigated here
$k=|l|+2n$, hence the poles are located at
$\lambda_{nl}=-i(|l|+2n+1)$ and
\begin{equation}
{\rm
Res}\,\<\phi|\Psi_{l\lambda}\>\Big|_{\lambda=\lambda_{nl}}=\frac{1}{\sqrt{(|l|+2n)!}}
\frac{\<\phi|\mathfrak{f}^-_{nl}\>}{\sqrt{2\pi}}\ ,
\end{equation}
where $\mathfrak{f}_{nl}^-$ is a distribution given by (\ref{f-nl}).

The conjugate distribution $\overline{\Psi_{l\lambda}}$ is defined
as
\begin{equation}\label{}
    \overline{\Psi_{l\lambda}}(\phi)=\<\phi|\overline{\Psi_{l\lambda}}\>=
    \frac1{2\pi}\int_{\mathbb{R}^2}e^{il\varphi}r^{i\lambda-1}
\overline{\phi}(r,\varphi)dS=\int_0^\infty
r^{i\lambda}\overline{\phi}_l(r)dr
\end{equation}
and it is regular in $\lambda=\lambda_{nl}$. The poles of
$\<\phi|\overline{\Psi}_{l\lambda}\>$ are located at
$\lambda=\overline{\lambda}_{nl}$. Hence
\begin{equation}\label{}
 \<\phi|\overline{\Psi_{l\lambda}}\>\Big|_{\lambda=\lambda_{nl}}=
 \sqrt{(|l|+2n)!}\,\frac{\<\phi|\overline{\mathfrak{f}^+_{nl}}\>}{\sqrt{2\pi}}\,,
\end{equation}
where $\mathfrak{f}^+_{nl}$ is a distribution given by
(\ref{f+nl}).

\section*{Appendix B}
\def\theequation{B.\arabic{equation}}
\setcounter{equation}{0}

Let us briefly proof that $\phi_l(r)$ given by (\ref{phi-n}) has a
power series expansion (\ref{phi-n_expansion}) starting from
$r^{|l|}$. Supposing that $\phi(x_1,x_2)$ is an analytic function
\begin{equation}\label{}
    \phi(x_1,x_2)=\phi(0,0)+\sum_{k_1,k_2}\frac{\partial^{k_1+k_2}\phi(0,0)}
    {\partial x_1^{k_1}\partial x_2^{k_2}}x_1^{k_1}x_2^{k_2}
\end{equation}
in cartesian coordinates $(x_1,x_2)$ it is obvious that in polar
$(r,\varphi)$-coordinates one obtains the following expansion for
$\phi_l(r)$:
\begin{eqnarray}
 \phi_l(r) &=&
 \frac1{2\pi}\int_0^{2\pi}e^{-il\varphi}\phi(r\cos\varphi,r\sin\varphi)\,d\varphi
 \nonumber  \\
 &=&\frac1{2\pi}\int_0^{2\pi}e^{-il\varphi}\Big[\phi(0,0)+\sum_{k_1,k_2}A_{k_1,k_2}r^{k_1+k_2}
 (\cos\varphi)^{k_1}(\sin\varphi)^{k_2}\Big]\,d\varphi \nonumber
 \\
 &=&\frac1{2\pi}\sum_{k_1,k_2}A_{k_1,k_2}r^{k_1+k_2}\int_0^{2\pi}e^{-il\varphi}
 (\cos\varphi)^{k_1}(\sin\varphi)^{k_2}\,d\varphi\, ,\label{polar-co-expansion}
\end{eqnarray}
where
\[ A_{k_1,k_2} = \frac{\partial^{k_1+k_2}\phi(0,0)}
    {\partial x_1^{k_1}\partial x_2^{k_2}} \ , \]
stand for derivatives of $\phi(x_1,x_2)$ in $(0,0)$. Now, the
question is: for which values of $k\equiv k_1+k_2$ the sum in
(\ref{polar-co-expansion}) does not vanish? Clearly, it should be
\begin{equation}\label{ineq}
 \int_0^{2\pi}e^{-il\varphi}
 (\cos\varphi)^{k_1}(\sin\varphi)^{k_2}\,d\varphi \neq 0\,.
\end{equation}
Using the Newton expansions
\begin{eqnarray*}
  (\cos\varphi)^{k_1} & = & \Big(\frac12\Big)^{k_1}
  \Big(e^{i\varphi}+e^{-i\varphi}\Big)^{k_1}= \Big(\frac12\Big)^{k_1}
  \sum_{m_1=0}^{k_1}{k_1 \choose m_1} e^{im_1\varphi}e^{-i(k_1-m_1)\varphi}\ , \\
  (\sin\varphi)^{k_2} & = & \Big(\frac1{2i}\Big)^{k_2}
  \Big(e^{i\varphi}-e^{-i\varphi}\Big)^{k_1}= \Big(\frac1{2i}\Big)^{k_2}
  \sum_{m_2=0}^{k_2}{k_2 \choose m_2}(-1)^{k_2-m_2}
  e^{im_2\varphi}e^{-i(k_2-m_2)\varphi}\ ,
\end{eqnarray*}
one can rewrite (\ref{ineq}) as
\begin{equation}\label{ineq-expansion}
 \Big(\frac12\Big)^{k_1}\Big(\frac1{2i}\Big)^{k_2}  \sum_{m_1=0}^{k_1}\sum_{m_2=0}^{k_2}
 {k_1 \choose m_1}{k_2 \choose m_2}(-1)^{k_2-m_2}
  \int_0^{2\pi}e^{-i(l+k-2(m_1+m_2))\varphi}d\varphi
  \neq 0\ ,
\end{equation}
hence (\ref{ineq-expansion}) will not vanish iff
\begin{equation}\label{cond}
    l+k-2(m_1+m_2)=0\,.
\end{equation}
Clearly,
\begin{equation}\label{m1-m2}
0\leq m_1\leq k_1\,,\qquad 0\leq m_2\leq k_2\,.
\end{equation}

Now, let $l<0$, so $l=-|l|$ and
\begin{equation}\label{cond2}
k=|l|+2(m_1+m_2)=|l|+2n\,,
\end{equation}
where $n=m_1+m_2\geq 0$. Due to (\ref{m1-m2}), in order to satisfy
(\ref{cond2}) it should be $k\geq |l|$.

On the other hand, if $l>0$, then
\begin{equation}
    k = -l +2(m_1+m_2)\leq -l+2k\,,
\end{equation}
because of (\ref{m1-m2}) and finally $k\geq l=|l|$. Note that in
this case $k=l+2(m_1+m_2-l)$, where $m_1+m_2-l\equiv n\geq 0$.

As a result we obtained that for a given $n$ the lowest power of
$r$ in expansion (\ref{polar-co-expansion}) is $k_1+k_2=|l|$.
Moreover, in both cases
\begin{equation}\label{}
    k_1+k_2=k=|l|+2n\,,\qquad n=0,1,2,\ldots
\end{equation}
holds.

\section*{Acknowledgments}

 This work was partially supported by the Polish State
Committee for Scientific Research Grant {\em Informatyka i
in\.zynieria kwantowa} No PBZ-Min-008/P03/03.

\end{document}